\documentstyle[12pt]{article}
\date{\ }
\begin{document}

\baselineskip=24pt

\title{Random Spin-$1$ Quantum Chains}
\author{Beatriz Boechat and Andreia Saguia \\
Instituto de   Fisica, Universidade Federal Fluminense \\ 
Campus da Praia Vermelha,  Niteroi, 24210-340, RJ, Brasil \\
Mucio A.Continentino \\
Instituto de Fisica, Universidade Federal Fluminense \\
Campus da Praia Vermelha, Niteroi, 24020-340, Rj, Brasil \\
and \\
National High Magnetic Field Laboratory \\
Florida State University, 1800 E.Paul Dirac Dr. \\
Tallahassee, Fl 32306 USA}
\maketitle
\begin{abstract}
\baselineskip=24pt
We study disordered spin-$1$ quantum chains with random exchange and 
biquadratic interactions using a real space renormalization group approach.
We find that the dimerized phase of the pure biquadratic
model is unstable and gives rise to
a random singlet phase in the presence of weak disorder.
In the Haldane region of the phase diagram we  obtain  a quite different
behavior.
\end{abstract}

\newpage

Although much progress has been made on the understanding of the
physical properties of ordered  isotropic
quantum spin chains, for both integer and half-integer spins \cite{affleck},
much
less has been achieved in the random case in particular for spin-$1$ chains.
For the spin-$1/2$ random exchange  Heisenberg  antiferromagnetic chain (REHAC)
theoretical
advance \cite{ma,fisher} has been motivated by experimental results mainly in
organic materials \cite{clark}.
The theory in this case gives a good description of the physical properties of
these systems which are governed by low energy excitations \cite{fernandes}. 
Also recently an extensive study using a renormalization group approach has
led
to a deeper understanding of the zero temperature phase diagram of
general spin-$1/2$ disordered quantum chains \cite{fisher}. 
On the other hand  the case
of  random, isotropic,  spin-$1$ chains has only recently been addressed
both experimentally \cite{aeppli,mucio1} and theoretically \cite{mucio1}.

In this Communication  we study a one-dimensional, 
spin-$1$,  Heisenberg system with random exchange and biquadratic
interactions.
We use a real space renormalization group approach which 
has proved to be very sucessful for spin-$1/2$
quantum chains \cite{ma,fisher}. We discuss the ground state properties and
extend 
the approach to finite temperatures to obtain the 
thermodynamic properties.

We consider the general bilinear, biquadratic, spin-$1$ chain with $L$ spins
described by the Hamiltonian
\begin{equation}
H = \sum_{r=1}^{L-1} J_{r} {\vec{S}}_r .  {\vec{S}}_{r+1} - 
\sum_{r=1}^{L-1} D_{r} \left(  {\vec{S}}_r  .  {\vec{S}}_{r+1} \right)^2
\end{equation}
where $J_{r}$ and $D_{r}$ are random nearest-neighbor
interactions with probability distributions $P_J(J_{r})$
and $P_D(D_{r})$ respectively and such that $0 \le J_{r} \le J$ and $0 \le
D_{r} \le D$.
The ${\vec{S}}_r$ are spin-$1$ operators and we are interested in the limit $L
\rightarrow \infty$.

We study in this Communication  the effect of disorder in two  distinct
situations:
\begin{enumerate}
\item The case where the biquadratic is the dominant interaction such 
that the cutoff
$D >> J$. We recall that the ordered system with  $D_{r} = D$ and furthermore
$J_{r} = 0$
corresponds to the so called $KBB$ point of the general phase diagram of the
Hamiltonian given by Eq.1 \cite{kbb}. At this point the purely biquadratic
Hamiltonian
can be solved exactly yielding a dimerized phase characterized by a two-fold
degenerate ground state with a {\em finite gap \/} to the excited states.

\item The cutoff $J$ of the exchange interaction is much larger than that of
the
biquadratic couplings, i.e. $J >> D$. This corresponds to investigate the
effect
of disorder on the section of the phase diagram of the pure system where a
Haldane gap occurs \cite{affleck}. The ground state of the ordered system ($J_r
= J$, $D_r = 0$)
in this case is unique and also has a gap for excitations.

\end{enumerate}

In order to deal with the Hamiltonian, Eq.1, we generalize, for the case
of spin-$1$  and biquadratic interactions,
a real space renormalization group method introduced by Ma, Dasgupta and Hu
\cite{ma}
to treat the spin-$1/2$ REHAC. The usefulness and limitations of this method
have been intensively discussed by Fisher \cite{fisher}
in recent papers on quantum random spin-$1/2$ chains.
In at least one of the situations we study  here this approach converges
to the exact solution of the random problem.
The method consists of  identifying the pair of spins with the strongest
coupling in the random chain and eliminating it by considering the
interaction with the neighboring spins of this pair as a perturbation.
This procedure generates  new couplings $J^{\prime}$ and $D^{\prime}$
between the two spins neighboring the eliminated pair and consequently
the form of the distribution $P_J(J_{r})$
and $P_D(D_{r})$ are modified.
One also gains a contribution to the
free energy of the system from the eliminated pair.
This process is then successively applied until a single pair of spins
remains.

We start at zero temperature and consider the case of purely biquadratic
interactions and no exchange. In the ordered case as mentionned above this
corresponds to
the so-called $KBB$ point \cite{kbb} of the general phase diagram
of the spin-1 chain given by the Hamiltonian of Eq.1.
We assume a distribution of random biquadratic couplings and initially take
$J_{r} = 0$.
Generalizing the procedure of  Ma et al. \cite{ma} for this problem,  we
identify the pair of spins with the strongest biquadratic coupling. When we
apply
the elimination transformation to this strong coupled pair we find an effective
coupling
between the spins neighboring the eliminated pair which is
is given by:
\begin{equation}
D^{\prime} = \frac{2}{9} \frac{D_1 D_2}{D}
\end{equation}
where $D_1$ and $D_2$ are the bonds connecting the strongest coupled pair of
spins,
with interaction strength $D$, to their neighbors.
It is clear that in this case the successive application of the elimination
transformation generates weaker and weaker couplings and the distribution of
biquadratic interactions becomes peaked close to the origin as the cutoff $D$
decreases.  The physical properties of the system are then dominated by low
energy
excitations and this new phase is very similar to
the {\em random singlet phase } of the disordered spin-$1/2$ random
antiferromagnetic chain \cite{fisher}. Consequently disorder has a dramatic
effect
at the $KBB$ point since in the pure case the system has a dimerized
phase with a gap for excitations and a short correlation length. The
introduction of weak
randomness on the biquadratic couplings gives rise to a new disordered phase
with low energy excitations and long-range magnetic correlations.
This behavior is universal in the sense that independently of the
form of the initial distribution of couplings $P_D(D_r)$ this converges to a
fixed point
distribution, which is approximately described by a power law
with a singularity at the origin \cite{ma}.
Notice that the equation we find here renormalizes to weak coupling
faster than in case of the spin-1/2 chain where the pre-factor of the
recursion relation  equivalent to Eq.2 is $1/2$ \cite{ma}.
We have found here a rather  unusual situation where a phase with a gap
and a short correlation
length has become completely unstable and furthermore with
long-range magnetic correlations due to the introduction of
arbitrarily weak disorder \cite{hyman}.

We include now a weak random exchange interaction between  nearest
neighbor spins such that the cutoff $J << D$.
A generalization of the  elimination transformation gives, for the new
exchange and biquadratic couplings  between
the spins neighboring the eliminated pair,  the following coupled relations
\begin{equation}
D^{\prime} = \frac{2}{9} \frac{D_1 D_2}{D+J_0}
\end{equation}
$$J^{\prime} = \frac{4}{3} \frac{J_1 J_2 }{\left( 3D+J_0 \right)}$$
where as before $D_1$, $D_2$, $J_1$ and $J_2$ are the bonds connecting
the strongly coupled pair of spins, with interactions $D$ and $J_0$ between
them, to their neighbors. Since $D >> J$ we expect, as before, that the
renormalized distributions converge rapidly to a weak coupling situation, i.e.,
to develop more and more weight close to the origin as the cutoff decreases.
Again the system attains a  random singlet phase dominated by low energy
excitations.

We remark that the recursion relations for the
biquadratic interactions iterate to zero faster
than that of the exchange couplings.  It may happen  that, depending
on the starting distributions,  the physics at sufficiently low energies
will be dominated by the exchange interactions. In this case
it is convenient to define the strongest coupled pair as
that for which $3D + J$, the energy to create an excited
state for a pair, is larger.

For completeness we give now the expression for the energy
associated with the eliminated pair and from which we can compute the
total ground state energy
$$E^{\prime} = -2J_0 -4D - \frac{4}{3} (D_1 + D_2) - \frac{4}{3} \frac{J_1^2 +
J_2^2}{J_0 + 3D}$$
\begin{equation}
- \frac{8}{27} \frac{D_1^2 + D_2^2 - D_1 D_2}{J_0 + D}
\end{equation}

We conclude this analysis of
case 1, i.e. $D >> J$,  pointing out that we have obtained
a new disordered phase for the random biquadratic spin-$1$
Hamiltonian which is similar to the random singlet phase of the spin-$1/2$
REHAC.

Case 2) We start with the situation where $D_r =0$.
For the random exchange spin-$1$ chain, the elimination
procedure yields the following relation for the new coupling
between the spins neighboring the eliminated pair
\begin{equation}
J^{\prime} =\frac{4}{3} \frac{J_1 J_2}{J}
\end{equation}
We find here quite a distinct situation from the previous case,
since this recursion relation appears to iterate to strong coupling.
The generation of couplings which are larger than those eliminated would in
fact
invalidate the present approach based on perturbation theory.
Contrary to the previous cases,  now the results depend strongly on
the form of the starting distributions.
When we include the biquadratic interaction, for $D << J$, the
new recursion relations are similar to those obtained before, namely
\begin{equation}
J^{\prime} = \frac{4}{3} \frac{J_1 J_2 }{\left( J + 3D_0 \right)}
\end{equation}
$$D^{\prime} = \frac{2}{9} \frac{D_1 D_2}{J + D_0}$$
for $D << J$. It is clear that the biquadratic is an irrelevant interaction,
in the renormalization group sense, for the random exchange spin-$1$ chain
since the starting  distribution $P_D(D_r)$ iterates to a distribution with
most of it`s weight at small coupling much faster than the exchange
distribution.
We notice that the biquadratic interactions reduce the new exchange couplings
generated by the elimination transformation. Whether this is sufficient to
guarantee the validity of the present approach will depend on the nature of
the starting distributions and the values of the cutoffs.

In order to get a better understanding of the random
exchange spin-$1$ chain, we have generalized the elimination
transformation procedure to finite temperatures \cite{mucio1}. In the
case $D_r = 0$ we get after some lenghty  calculations
\begin{equation}
J^{\prime} = \frac{4}{3} \frac{J_1 J_2}{J} W(\beta J)
\end{equation}
where $J_1$ and $J_2$  as before are the bonds connecting
the strongly coupled pair of spins to their neighbors. We point
out that this transformation does not generate new terms preserving
the form of the original Heisenberg Hamiltonian.
The free energy has a contribution from the eliminated pair which turns out to
be,
\begin{equation}
F^{\prime} = F_0 - \frac{4}{3} \frac{J_1^2 + J_2^2}{J} V(\beta J)
\end{equation}
with
\begin{equation}
F_0 = -2J - \frac{1}{\beta}  Ln(1+3e^{-\beta J}+5e^{-3 \beta J)})
\end{equation}
The quantities W(y) and V(y) are given by \cite{mucio1}:
\begin{equation}
W(y)= \frac{1-(3/8)e^{-y}(1+y) - (5/8)e^{-3y}(1+3y)}
{1+3e^{-y} + 5e^{-3y}}
\end{equation}
and
\begin{equation}
V(y) = \frac{1-(3/8)e^{-y}(1-y) - (5/8)e^{-3y}(1-3y)}
{1+3e^{-y} + 5e^{-3y}}
\end{equation}
Notice that for $k_B T > 0.37J$ the function  $W(\beta J) <(3/4)$ and
this guarantees the validity of the present approach for
sufficiently high temperatures independent of the starting distribution.

For certain probability distributions, which we associate with 
the case of {\em strong disorder \/}, the 
successive elimination transformations
give rise to weaker and weaker couplings as the 
cutoff $J$ decreases. In this case, for sufficiently
small $J$, the distribution
$P_J(J_r)$ becomes peaked at $J_r \approx 0$ and can be  approximated
by a power law \cite{ma,mucio1},
\begin{equation}
P_J(x,J) \approx  \frac{\alpha}{J} x^{-1+ \alpha}
\end{equation}
where $x=J_r/J$, $J$ is the actual cutoff and the factor
$(\alpha / J)$ arises from the normalization condition.
This  distribution leads to a low temperature
behavior of the thermodynamic 
quantities which can be described by power laws \cite{ma,mucio1},
\begin{equation}
F \propto T^{1+ 2 \alpha},   C_v = -T \frac{{\partial}^2 F}{\partial F^2}
\propto T^{\gamma_c}
\end{equation}
and
\begin{equation}
\chi \propto T^{-1+ \gamma_s},   m \propto H^{\gamma_h}
\end{equation}
where $\gamma_c$ is the specific heat exponent, 
$\gamma_s$ and $\gamma_h$ are
related to the exponent $\alpha$ of P(x,J) 
and of the free energy $F$.
The susceptibility exponent is given, to a good approximation by, 
$\gamma_s \approx 2 \alpha$ and furthermore
$\gamma_h \approx \gamma_s$ \cite{ma}. $H$ and $m$ are respectively
the uniform external field and magnetization. Since the power law distribution
is not
the exact fixed point distribution for the elimination transformation,
$\alpha$ and consequently 
$\gamma_c, \gamma_s$ and $\gamma_h$
depend on temperature and cutoff.
For initial power law distributions, Eq.12, this dependence
is very weak \cite{ma,mucio1}.

The results contained in Eqs.13 have been explicitly verified for
the {\em strong\-ly disordered}
cases of 
initial
uniform, $P(K)= \Theta(1-K)$,
and power law distributions (Eq.12) \cite{mucio1}.
The free energy was  obtained numerically from
the accumulation point
of Eq.8 for successive elimination transformations. Then  Eqs.13 yield
the exponent $\alpha$ and the specific heat \cite{mucio1}.
For the same starting distribution and temperature range  we found that the
exponent $\alpha$
for the spin-$1$ chain is consistently
larger than that for the spin-$1/2$.  In fact,
for any initial
distribution which attains
a power law form whenever the cutoff is sufficiently
small, we can show that the exponent $\alpha$ decreases more rapidly when
the cutoff is further reduced for $S=1/2$ than for $S=1$ \cite{mucio1}.
These theoretical results are in agreement with experiments in the strongly
disordered, one-dimensional, spin-1 system $MgVOBO_3$ which shows
power law behavior of the susceptibility in  a wide temperature
range \cite{mucio1} although less singular than in the isostructural
spin-1/2 material $MgTiOBO_3$ \cite{fernandes}.

The elimination transformation 
for $S=1$,  Eq.7 or Eq.5, applies for {\em 
strongly disordered chains \/}, which we generically define as those
characterized by initial distributions which yield thermodynamic properties
at low temperatures as described by Eqs.13 and 14.
For these distributions couplings larger than the one eliminated are
generated with negligeable probability and the cutoff decreases
rapidly  enough  to yield meaningful thermodynamic behavior as in the
cases of uniform and power law distributions. Otherwise
our results represent a good approximation
for the high temperature behavior of disordered spin-1 chains.

We have extended the elimination transformation method for a quantum Heisenberg
chain with arbitrary spin. We find the following recursion relation
\begin{equation}
J^{\prime}= \frac{2}{3} S(S+1) \frac{J_1 J_2}{J}
\end{equation}
This result shows that perturbation theory fails for  large spins
since the excited states become nearly degenerate with the ground state.
The critical value of $S$ is between $S=1/2$ and $S=1$ so that a universal
random singlet type of phase in the Heisenberg chain with exchange interactions
only is guaranteed  for spin-1/2. In this case the quantum fluctuations are
sufficiently strong and the renormalization group equations
yield a flow to a fixed point universal distribution singular at the origin.

In conclusion we have generalized an elimination transformation procedure to
study the
effect of disorder on quantum antiferromagnetic chains with exchange
and biquadratic interactions. In case the biquadratic interaction dominates,
we find a new disordered phase for spin-$1$ chains which is essentially a
random singlet phase.
We find that weak disorder destroys the dimerized phase of the ordered system
giving rise to
low energy excitations and long range correlations.
In the Haldane region of the phase diagram of the pure chain our results
indicate that
weak disorder has not a profound effect although the present approach
fails in this  case at least at low temperatures. However for starting
distribuitons
which are strongly disordered we obtain a
thermodynamic behavior which is similar to the random singlet
phase of disordered spin-$1/2$ REHAC but with weaker singularities.

\begin{center}
\bf Acknowledgements
\end{center}
We would like to thank Prof. E. Muller-Hartmann and Prof. J.C.Fernandes for
invaluable discussions.

\end{document}